# Gravitational Microlensing by Clustered MACHOs

**R. Benton Metcalf and Joseph Silk**
Department of Physics and Department of Astronomy
University of California, Berkeley, CA 94720
e-mail: bmetcalf@pac2.berkeley.edu

**Abstract**

It has been proposed that the MACHOs in our galaxy could be clumped in globular cluster–like associations or RAMBOs (robust associations of massive baryonic objects) (Moore & Silk 1995). Here we investigate the effect such clustering has on the microlensing of stars in the Large Magellanic Cloud. We find that the lensing in a one square degree field could be dominated by just a few clusters. As a result the lensing properties vary widely depending on the position and velocity of those clusters which happen to lie between us and the LMC. Moreover we find a large variance in timescale distributions that suggests that the small number statistics could easily be dominated by events in the tails of the unclustered distribution (e.g. by long periods). We compare our results with the MACHO collaboration's data and find that a "standard" halo made entirely of MACHOs is not strongly disfavored if the clusters have masses of $10^6 M_\odot$. For less massive clusters such a halo is not as likely. For $10^4 M_\odot$ clusters the microlensing statistics are essentially unchanged from the unclustered case. It may be possible to detect very massive clusters from the distribution of events in time-scale and space. We provide some example time-scale distributions.

## 1. Introduction

Baryonic dark matter (BDM) is an attractive candidate for halo dark matter. BDM is known to exist, is plausibly produced in the vicinity of luminous matter with which it shares a common origin, and primordial nucleosynthesis constraints suggest an amount of BDM that is approximately equivalent to that required for halo dark matter. Since diffuse halo gas cannot account for the dark matter, compact halo objects of substellar mass (MACHOs) provide the most likely candidates for BDM. Dark stellar mass objects, such as white dwarfs, neutron stars or black holes, cannot definitively be excluded as halo dark matter, but the luminosity produced during their formation and their nucleosynthetic byproducts severely constrains such interpretations to a small subset of parameter space (*e.g.* Charlot and Silk 1995).

Two gravitational microlensing experiments (MACHO and EROS) have independently discovered events which were most likely the result of gravitational microlensing by MACHOs. The most complete study of detection efficiency, performed for the MACHO experiment, concludes that in the simplest halo model, taken to be uniform, spherical and isotropic in velocity dispersion, only $\sim 20\ \%$ of halo dark matter can be in MACHOs over the mass range $10^{-6}$–$10^{-1} M_\odot$. The 95 % confidence upper limit is $\sim 50\ \%$ of the halo mass consisting of MACHOs in this mass range.

Considerations of galactic star formation suggest that the predominant mode of star formation is to form stars not in isolation but in clusters. Star formation was certainly different during the formation of the halo in order to have formed even the observed population of MACHOs, which greatly dominate ordinary halo stars in terms of halo mass fraction. Why did the characteristic stellar mass during halo formation peak in the range $10^{-2} - 10^{-1} M_\odot$, as inferred if we take the microlensing results towards the LMC at face value?

One could speculate that the lack of a magnetic field in pregalactic halo clouds may have inhibited angular momentum transfer, thereby only allowing low mass fragments to form and contract. Theoretical reasons apart, it is clear however that a BDM–dominated halo and the MACHO detections together require a bottom-heavy IMF, one whose characteristic mass is at least an order of magnitude or more below that of the solar neighborhood IMF.

One can say a little more about the expected masses of the MACHO clusters. Cosmological considerations of the minimum Jeans mass suggest a typical pregalactic mass scale of $10^4 - 10^6 M_\odot$. Studies of thermal instability in the protogalaxy, as a preliminary to globular cluster formation, indicate a typical mass scale, based on pressure equilibrium with virialized halo gas, of $\sim (T_{cold}/T_{hot})^2 M_B$, where $M_B$ is the galactic mass in baryons, $T_{cold}$ is the temperature of cooled gas, and $T_{hot}$ is the temperature of virialized gas (Fall and Rees 1985). With $T_{cold} \approx 10^4$ K, as expected from primordial atomic gas cooling, and $T_{hot} \approx 3 \times 10^7$ K, from the gravitational equilibrium of diffuse halo gas, one infers a typical cloud mass of $\sim 10^6 M_\odot$. Analogous reasoning suggests that if some molecular cooling occurs, via formation of $H_2$, $HD$, and $LiD$ molecules, the cold phase attains a temperature of $\sim 100$ K. In this case, typical clump masses may be as small as $\sim 100 M_\odot$. Evidently, one should consider the possibility of a wide range of clumpiness, spanning the range $\sim 10^2 - 10^6 M_\odot$, for the characteristic scale of the parent clouds of halo MACHOs.

MACHO clusters would be dynamically stable in the outer halo, where tidal interactions are unimportant. In the inner halo, dynamical arguments limit the possible $(M, R)$ parameter space to a relatively narrow range, defined by tidal disruption in the galactic gravitational potential and by disruption of globular star clusters:

$$10 r_{pc}^3 M_\odot \lesssim M \lesssim 1000 r_{pc}^2 M_\odot; \quad r_{pc} \equiv (r/1pc) \lesssim 70. \tag{1}$$

In this paper, we study how MACHO clumpiness can affect experimental determinations of the halo dark mass fraction, and describe how discreteness in event statistics may eventually allow detection of MACHO clustering.

## 2. Microlensing by MACHO clusters

We first describe the techniques that we use to estimate the microlensing properties of a halo consisting of clustered MACHOs. The average density of clusters at position $x$ is $\rho(x)/M_c$. For the mass density of the halo we follow many other authors by using a simple spherical model with a core,

$$\rho(r) = \rho_0 \frac{r_0^2 + a^2}{r^2 + a^2}. \tag{2}$$

The density at the solar radius, $r_0$, is $\rho_0$ and $a$ is the core radius. With characteristic values $\rho_0 = 7.9 \times 10^3 M_\odot pc^{-3}$, $a = 5$ kpc and $r_0 = 8.5$ kpc, there are an average of 10 to 800 clusters intersecting a one square degree field in the LMC for cluster masses $10^6 M_\odot$ to $10^4 M_\odot$. Because the numbers are so small the microlensing properties can vary widely depending on the particular positions and velocities of the few clusters that are responsible for most of the microlensing. For this reason, it is better to take the approach of calculating the microlensing properties for particular realizations of the cluster positions and velocities rather than averaging over these realizations and thereby washing out distinctive properties. Taking a particular realization of the cluster positions and velocities is justifiable because these quantities are essentially static over the duration of any conceivable observating schedule.

With the above in mind we will in general calculate the microlensing properties of one cluster and then add up the contributions of each cluster. The clusters are taken to be spheres of radius $r_c$ and mass $M_c$ with MACHOs of mass $m$ uniformly distributed within them. The distance to the source stars is $D_s$. For the lensing of LMC stars it is a good approximation to take all the source stars to be at the same distance. Take the center of a cluster to be located a distance $xD_s$ from the observer in the radial direction. The position perpendicular to the radial direction will be denoted by $x_\perp$. The optical depth is the probability per source star of a MACHO being within one Einstein ring radius, $R_e$, of the line of sight connecting us to any one of the source stars. The volume within $R_e$ of the line of sight is often called the microlensing tube. Since the dimensions of the clusters are small in comparison to the distance to the LMC it is a good approximation take $R_e$ to be constant within the volume of the cluster for a given MACHO mass, m. If the source stars are evenly distributed across observed field the optical depth of one cluster is then the fraction of the volume at $xD_s$ that is within the lensing tubes, times the volume of the cluster within the field, times the number density of MACHOs within the cluster, divided by the number of source stars. This works out to

$$\tau_c = \frac{M_c \pi R_e^2}{m \Omega D_s^2 x^2} f(r_c, x_\perp, xD_s, \Omega) \tag{3}$$

where $\Omega$ is the solid angle being observed and $f(r_c, x_\perp, xD_s, \Omega)$ is the fraction of the cluster that lies within that solid angle.

Since the velocity dispersion of MACHOs within a cluster is expected to be small in comparison to the net velocity of the cluster, $\sigma_{int} \approx \sqrt{GM_c/4r_c}$, we will neglect it (Moore & Silk 1995). Griest (1991) showed that the velocities of the sun and the LMC contribute relatively little to the microlensing statistics so we ignore them. It was realized by Paczyński (1986) that in this case the microlensing rate, $\Gamma$, of events of time-scale, $t_o$, is equal to the inverse of the mean time between lensing events of that time-scale. So for each cluster

$$d\Gamma_c = <\Delta t>^{-1} = \frac{2\tau_c V_c}{\pi R_e}\frac{dn}{dm}dm = \frac{2\tau_c}{\pi t_o}P(t_o)dt_o \qquad (4)$$

where $\frac{dn}{dm}$ is the normalized mass function of MACHOs and $P(t_o)$ is the "static" probability distribution of time-scales. This probability is given by

$$P(t_o)dt_o = \frac{t_o(cV_c)^2}{2GD_sx(1-x)}\frac{dn}{dm}\left(m = \frac{(cV_c t_o)^2}{4GD_sx(1-x)}\right)dt_o. \qquad (5)$$

To find the rate from this it would be more convenient to integrate over $m$ rather than $t_o$ were it not for the observational efficiency which is a function of time-scale. In the case of a continuous distribution of MACHOs the variables $x$ and $V$ must also be integrated over with the velocity distribution. For clustered MACHOs we just add up the $\Gamma_c$'s. It is shown in the appendix that within the regime of interest the optical depth and lensing rate are equal to those of the unclustered case if they are averaged over all configurations of the clusters.

## 3. Results

Microlensing properties were calculated by first calculating the average number of clusters that should lie within the volume in which they intersect the field of view. This is done by integrating over the average halo density given in 2. The field of view is taken to be circular for simplicity of calculation. Then the actual number of clusters within this volume is chosen assuming Poisson statistics. For each of the clusters a position within the volume is chosen at random from a probability distribution that is proportional to the mass density. The fraction of the cluster within the field of view is calculated using Monte Carlo integration and a perpendicular velocity for each cluster is chosen from the distribution,

$$f(v)d^2v = \frac{e^{-v^2/\sigma^2}}{\pi\sigma^2} \qquad (6)$$

where $\sigma = 220$km/s. With these quantities the microlensing characteristics of each cluster are calculated and added together to get the results for that realization.

The optical depths for cluster masses $10^6 M_\odot$, $10^5 M_\odot$ and $10^4 M_\odot$ were calculated for circular fields of one square degree. As expected, in each case the average optical depth is equal to the value it would have were the MACHOs unclustered, $\tau_{av} = 5.1 \times 10^{-7}$. The standard deviations in the optical depth are found to be $2.8 \times 10^{-7}$, $1.0 \times 10^{-7}$ and $0.3 \times 10^{-7}$ for cluster masses of $10^6 M_\odot$, $10^5 M_\odot$ and $10^4 M_\odot$ respectively. However the distributions for the larger mass clusters are not very symmetric. Their coefficients of skewness are $3 \times 10^{-7}$, $3 \times 10^{-8}$ and $5 \times 10^{-9}$ respectively.

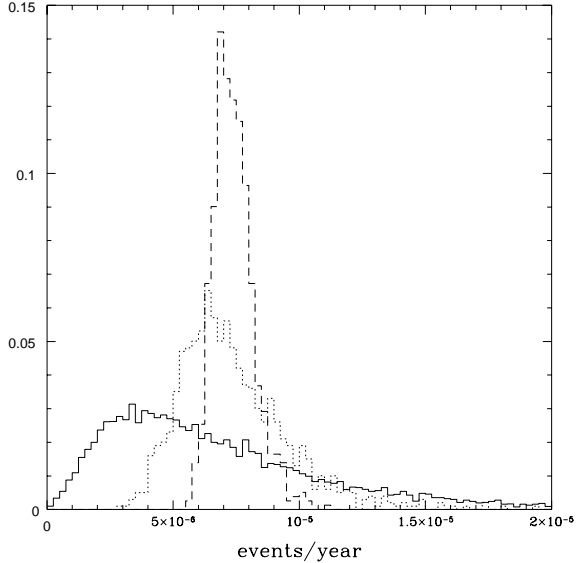

Figure 1: The distribution of rates amongst different realizations of cluster positions and velocities. The solid line is for $M_c = 10^6 M_\odot$, the dotted line is for $M_c = 10^5 M_\odot$ and the dashed line is for $M_c = 10^5 M_\odot$. All curves have the same arbitrary normalization and in all cases $<\sqrt{M_\odot/m}>= 4.47$.

The microlensing rates also have an average value that is in agreement with previous calculations for unclustered MACHOs (Griest 1991), $\Gamma_{av} = 1.7 \times 10^{-6}\sqrt{M_\odot/m}\,\text{yr}^{-1}$. The standard deviations are $1.4 \times 10^{-6}\text{yr}^{-1}$, $0.5 \times 10^{-6}\text{yr}^{-1}$ and $0.2 \times 10^{-7}\text{yr}^{-1}$ times $\sqrt{M_\odot/m}$. The distributions of rates are shown in figure 1. These rates and depths are calculated for a one square degree field so this distribution represents the expected fluctuations amongst fields of that size.

To calculate time-scale distributions we must assume a mass-function for the MACHOs. The mass function for normal hydrogen containing objects is constrained on the high side by searches for low-mass hydrogen burning stars (Hu *et al.* 1994, Bahcall *et al.* 1994). Below $m \sim 0.1 M_\odot$ the number density of halo objects is not appreciable constrained by conventional observations. We choose to approximate the mass function with a normal distribution

$$\frac{dn}{dm} \propto e^{(m-\bar{m})^2/\sigma_m^2}. \qquad (7)$$

The average mass is chosen to agree well with the microlensing data, $\bar{m} = 0.05 M_\odot$ and $\sigma_m = 0.01 M_\odot$. Representative time-scale distributions are plotted in figures 2 - 4 along with the time-scale distribution for unclustered MACHOs with a delta function mass function centered on $\bar{m}$. For the high mass clusters only a handful of clusters, $\sim 8$ are responsible for most of the microlensing so that the time-scale distribution can vary substantially from the unclustered case and from other possible realizations of cluster positions and velocities.

The microlensing data favors a very steeply peaked mass function (Alcock, et al. 1995b). A more peaked mass-function would result in steeper and more distinct peaks in the time-scale distribution. Time-scale distributions for $M_c = 10^4 M_\odot$ are unlikely to be distinguishable from no clusters without substantially more microlensing events than are likely in the near future.

**4. Comparison with Data**

In order to use the present observations to constrain our models we must take into account the efficiency in the observations. We use the MACHO collaboration's data and calculate the efficiency by interpolation from figure 2 of Alcock et al. (1995a). To find the expected number of events, $N_{ex}$, we calculate the time-scale for each cluster assuming a delta function mass- function and then multiply the rate for that cluster by the efficiency at that time-scale. This effective rate is multiplied by the number of observed stars , $8.6 \times 10^6$, and the period of time over which they were observed, 1.1 years. We approximate the observed fields as consisting of multiple one square degree circular fields.

To constrain the model parameters $m$ and the fraction of the halo density in MACHOs we use the baysian maximum likelihood approach. For each choice of these parameters the expected number of events, $N_{ex}$, depends on where the clusters happen to be located and at what velocities they happen to be traveling. In turn the numbers of events that will be observed are assumed to be Poisson distributed about the expected number of events. The likelihood function also contains factors that are proportional to the probability of observing events with time-scales equal to those observed. For the three events observed by the MACHO collaboration we use the likelihood function

$$\frac{1}{\Gamma_{eff}(m)^3} \frac{d\Gamma_{eff}(m,t_1)}{dt_o} \frac{d\Gamma_{eff}(m,t_2)}{dt_o} \frac{d\Gamma_{eff}(m,t_3)}{dt_o} \sum_{N_{ex}} N_{ex}^3 e^{-N_{ex}}. \qquad (8)$$

The summation is over all the realizations for a given $M_c$ discussed in section 3. If the MACHOs are unclustered the summation is unnecessary since in this case $m$ and the halo fraction predict a unique $N_{ex}$. $\Gamma_{eff}(m)$ is the rate adjusted for the efficiency. For simplicity we estimate $\Gamma_{eff}$ with its unclustered value. It should be noted that the factors outside the summation could fluctuate quite wildly depending on the particular configuration of clusters. This effect is essentially averaged out by taking the unclustered value for $\Gamma_{eff}$.

The observed time-scales, $t_1$, $t_2$ and $t_3$ are 19.4 days, 10.95 days and 15.6 days (Alcock 1995b). These time-scales have been adjusted for the effects of blending. Note that we use a definition for the time-scale that differs by a factor of two from the one used by the MACHO collaboration: crossing time over an Einstein ring radius rather than diameter. The likelihood plots for the "standard" halo model ( equation 2 ) with three values of the cluster mass are given in figures 5 - 7. Also shown are the likelihood contours for the unclustered case. Our likelihood plots differ somewhat from the plot given in Alcock *et al.*(1995a) because in that paper the *a priori* probability density in parameter space is taken to be $\propto dm\, df/m$ where $f$ is the halo fraction in MACHOs. We take this probability to be uniform, $\propto dm\, df$. This causes our likelihood limits to be somewhat more restrictive in the mass dimension and less restrictive on the halo fraction. We also take $D_s = 55$ kpc where as they use 50 kpc which has the opposite affect.

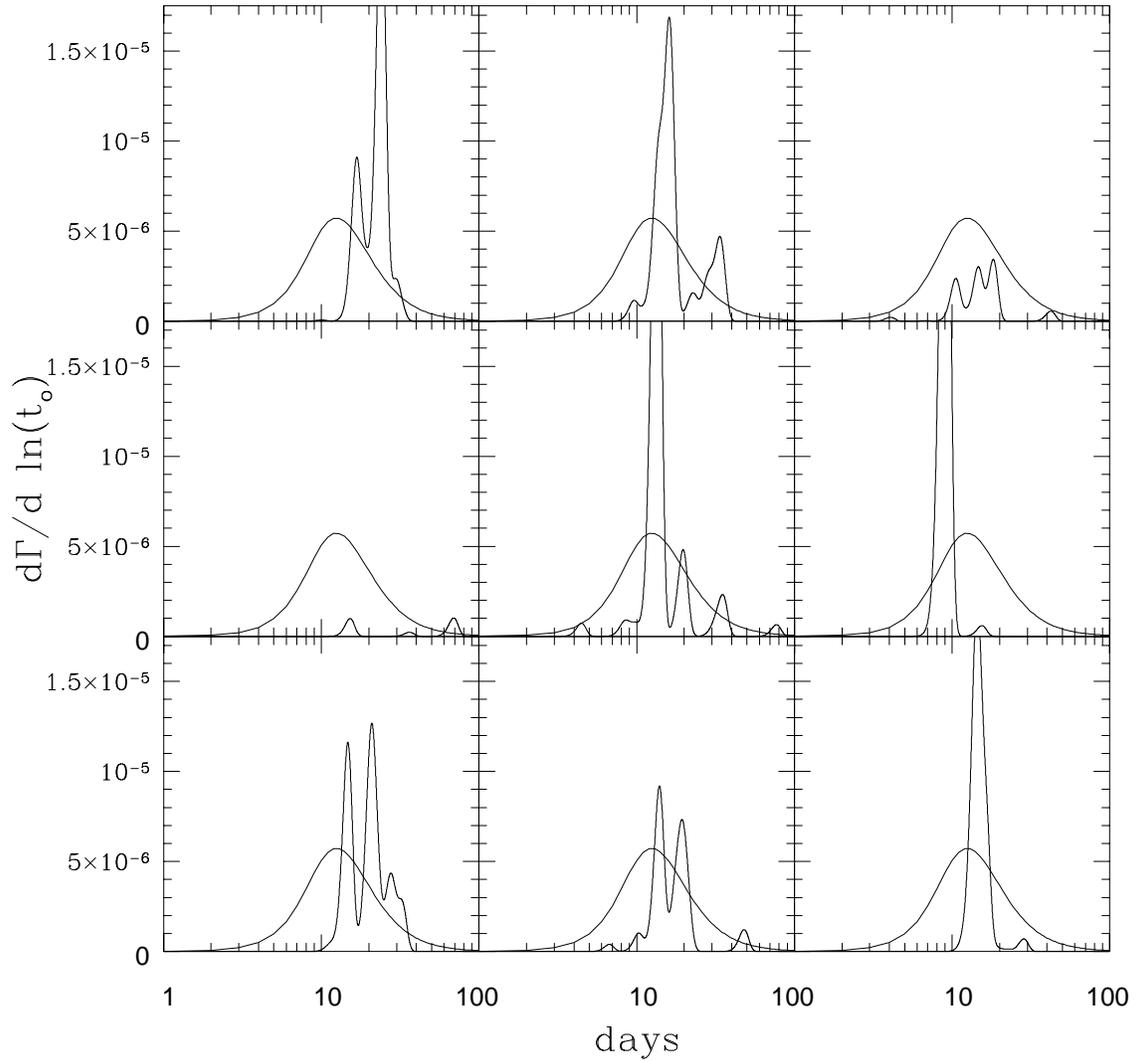

Figure 2: Sample event time-scale distributions for $10^6 M_\odot$ clusters. Each plot is a separate realization of cluster positions and velocities. The smooth curves are the time-scale distribution for the case of no clustering and $m = 0.05 M_\odot$.

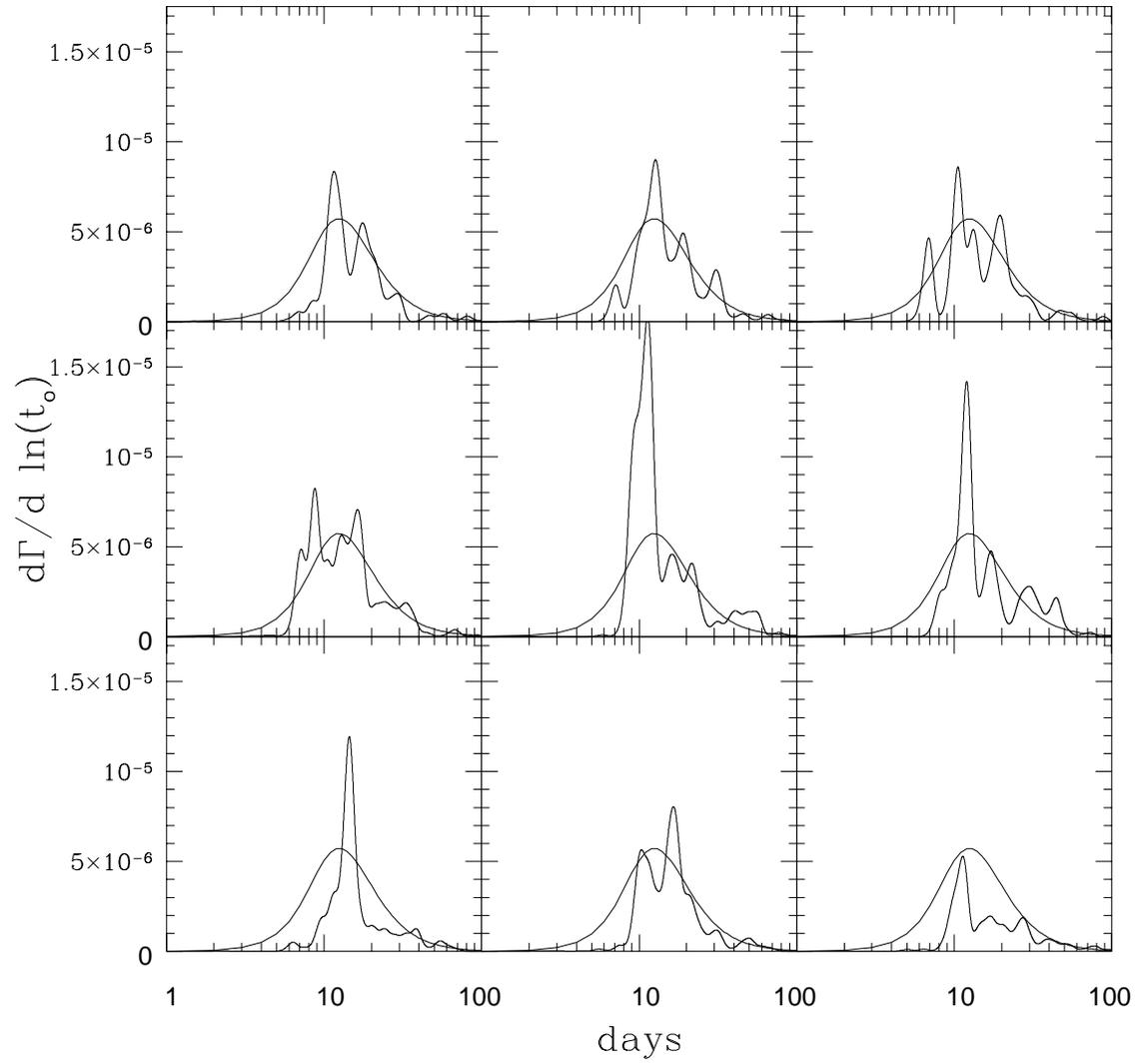

Figure 3: Sample event time-scale distributions for $10^5 M_\odot$ clusters.

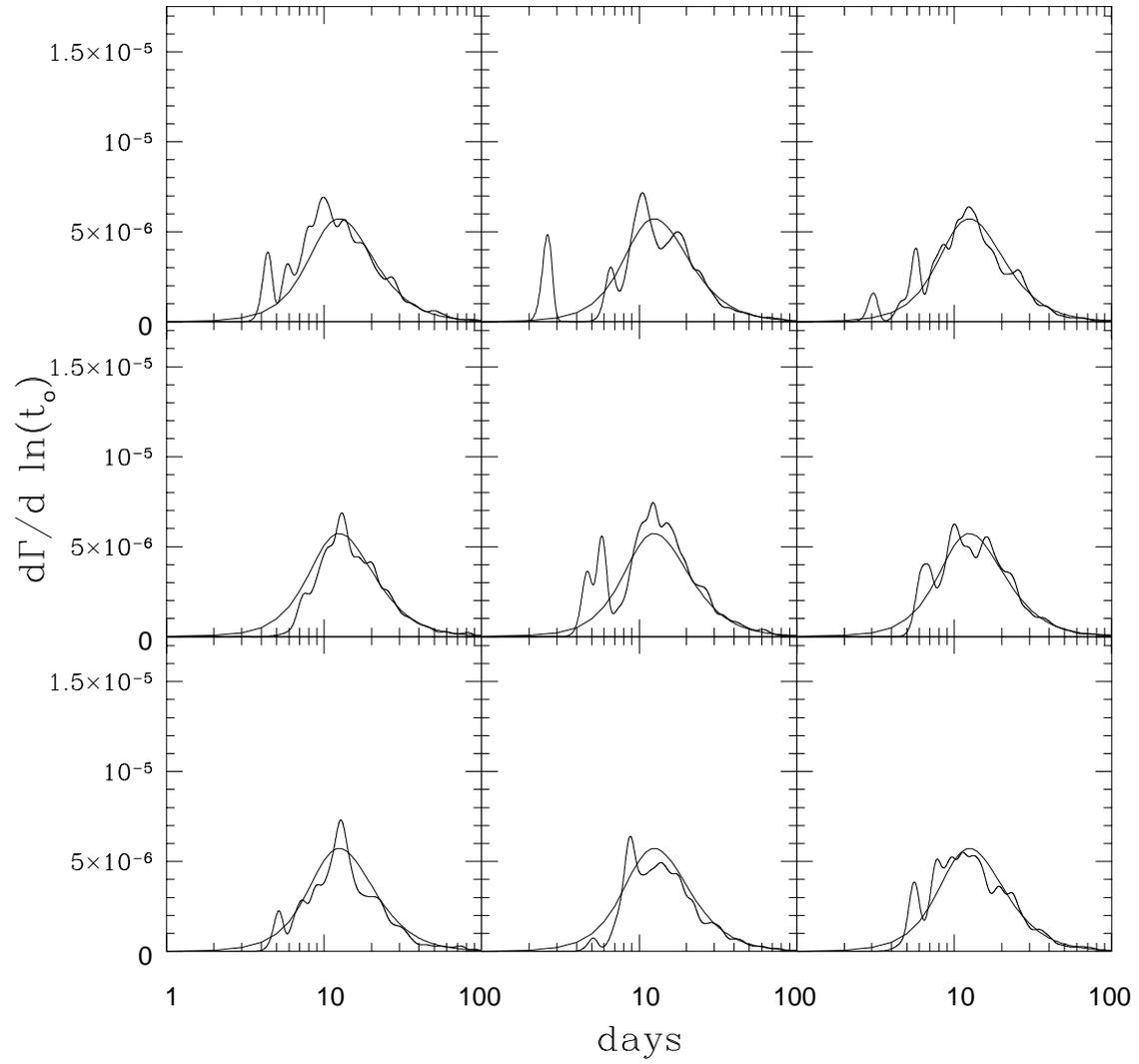

Figure 4: Sample event time-scale distributions for $10^4 M_\odot$ clusters.

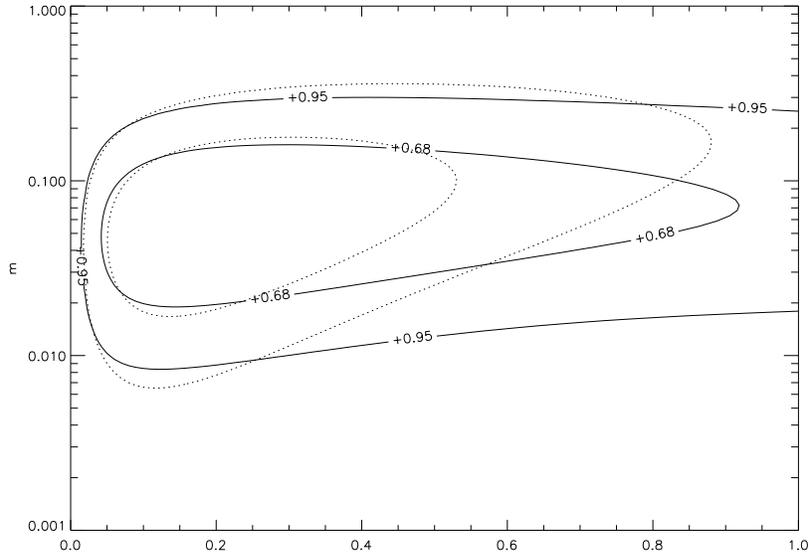

Figure 5: Likelihood plot for $10^6 M_\odot$ clusters given the MACHO collaboration's data. The vertical axis is the mass of the MACHOs and the horizontal axis is the fraction of the halo that is composed of MACHOs. The dotted contours are for the unclustered case with the same likelihood intervals.

We also repeated this whole procedure for a flattened halo model of the type devised by Evans (Evens 1994,Evens & Jijina 1994). The mass density is given by

$$\rho(R,z) = \frac{v_o^2 a^\beta}{4\pi G q^2} \frac{a^2(1+2q^2) + R^2(1-\beta q^2) + z^2\left(2 - \frac{1+\beta}{q^2}\right)}{(a^2 + R^2 + z^2/q^2)^{(4+\beta)/2}}. \qquad (9)$$

Here $(R,z)$ are cylindrical polar coordinates. Figures 8 and 9 give the likelihood plots for this model with the parameters $a = 5$ kpc, $v_o = 200$ km/s, $\beta = 0.0$ and $q = 0.7$. With these values the model has an asymptotically flat rotation curve and approximates an E6 in shape. The velocity distribution is modeled in the same way as before. This model predicts a greater amount of microlensing then the "standard" model does and thus the data sets a slightly stronger constraint on parameter space. As in the spherical case the constraints for $10^4 M_\odot$ clusters are virtually identical to the unclustered case.

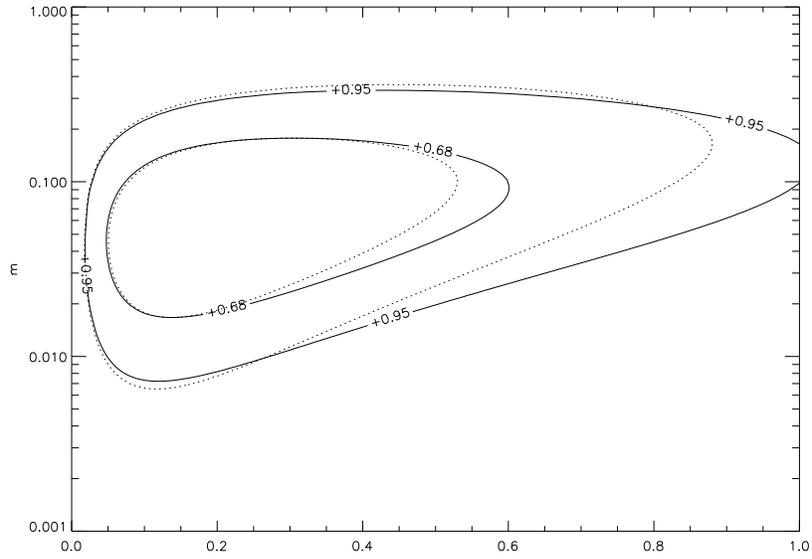

Figure 6: Likelihood plot for $10^5 M_\odot$ clusters.

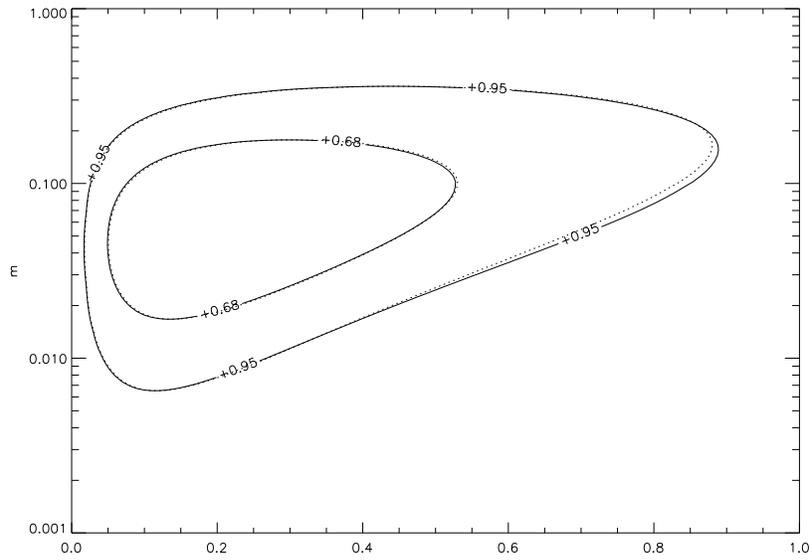

Figure 7: Likelihood plot for $10^4 M_\odot$ clusters.

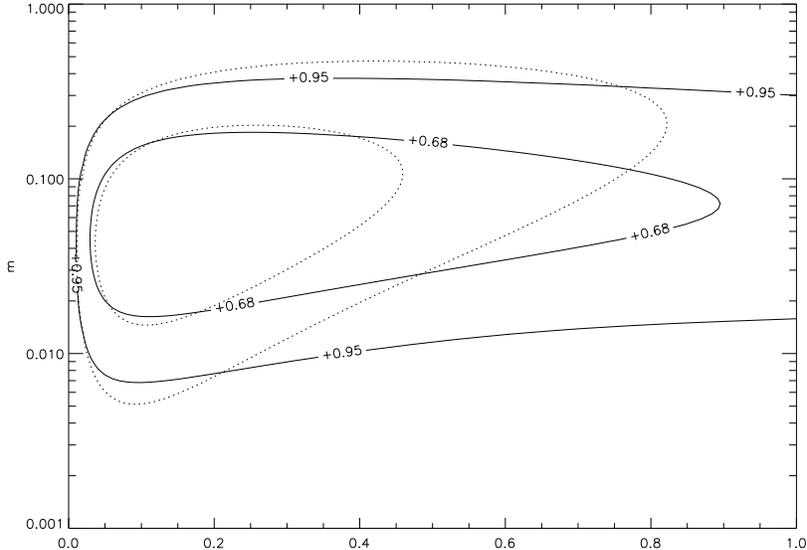

Figure 8: Likelihood plot for $10^6 M_\odot$ clusters in the flattened halo model.

In this Baysian analysis we have taken the liklihood-function to vanish outside the range of these plots. Because of this the MACHO fraction is less than one by convention. This can produce problems with interpretation. For instance if the liklihood-function is integrated over MACHO masses to get a one-dimensional liklihood-function for the MACHO fraction, the result will always exclude a 100% MACHO halo to an arbitrarily high confidence level. This makes little difference if the likelihood function is small on the borders of the allowed region of parameter space. This is not the case for our $10^6 M_\odot$ and $10^5 M_\odot$ cluster models. If the allowed region of parameter space is extended so that a MACHO fraction of 2 is possible, the the upper limits on the MACHO fraction change. The 95% confidence level upper bound on the $M_c = 10^5 M_\odot$ flattened model goes from 0.70 to 0.77. The effect is much more pronounced for $10^6 M_\odot$ clusters where the same limit goes from 0.9 to 1.7 and the 68% confidence limit goes from 0.5 to 0.8. In this sense, models with $10^6 M_\odot$ clusters are almost completely unconstrained by the microlensing data except in the MACHO mass

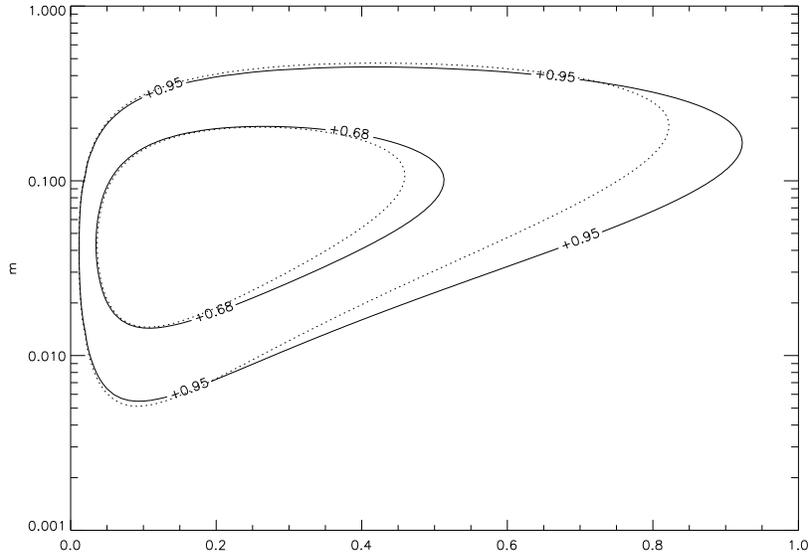

Figure 9: Likelihood plot for $10^5 M_\odot$ clusters in the flattened halo model.

domain.

## 5. Conclusions

For the models we have considered, a halo made entirely of MACHOs in $10^6 M_\odot$ clusters is fully consistent with the microlensing data, while $10^5 M_\odot$ MACHO clusters would constitute $\lesssim 80\%$ of the halo density. Many other halo models are possible. Alcock *et al.* (1995b) have pointed out that if the galactic disk is large, it is not necessary to have as massive a halo and the microlensing rate could be substantially smaller. Some of their 100% MACHO models are also consistent with the present data. Our findings loosen the constraint even more. One intersesting asspect of our results is that we find a large variance in timescale distributions for massive MACHO clusters. This suggests that the small number statistics could easily be dominated by events that would be in the tails of the unclustered distribution (e.g. by long periods).

However, the future looks bright. With more data towards the LMC and SMC, spatial variations in the microlensing rate will to some extent be measurable. This will constrain halo models as well as clustering. The level of clustering can also be measured in the time-scale domain. A number of special purpose statistics could be used to identify clustering in the time-scale distribution. An example is $\log(t_{max}/t_{min})/\log(t_i/t_{i+1})$ where $t_i$ and $t_{i+1}$ are the two time-scales that are closest together and $t_{max}$ and $t_{min}$ are the two furthest apart. The significance of this statistic is not strongly dependent on the halo model or models for LMC and disk lensing. Because of this it would serve as a good test of clustering irrespective of other parameters. Currently the value of this statistic is not statistically significant, but it is not expected to be until the number of events becomes substantially larger then the number of peaks in the true time-scale distribution.

**Appendix**

The correspondence between our optical depth and rate averaged over all cluster configurations and the depth and rate due to unclustered MACHOs follows directly from the fact that

$$\int d^2 x_\perp f(r_c, x_\perp, xD_s) \simeq A(xD_s) \tag{10}$$

to high accuracy. Here $A(xD_s)$ is the area enclosed within the angular field of view at a distance $xD_s$ from the observer. $A(xD_s)$ is equal to $\Omega x^2 D_s^2$ for the case of a circular field. To see that this is true, we can express the volume fraction as the ratio of the volume within the field of view, $\bar{V}_c$, to the total volume of the cluster, $V_c$. We then express $\bar{V}_c$ in terms of a integral over a function $\Theta(\bar{z})$ which is defined in cluster centered coordinates to be equal to unity within the cluster's volume and zero everywhere else.

$$\frac{1}{V_c} \int d^2 x_\perp \bar{V}_c(x_\perp) = \frac{1}{V_c} \int d^2 x_\perp \int_v d^3 z \Theta(z - x_\perp). \tag{11}$$

The $z$ integral is over the volume within the field of view. We now switch the order of integration and realize that $\Theta(z - x_\perp)$ integrated over $d^2 x_\perp$ is just the area of a slice of the cluster at distance $z_3$ form the observer. Call this area $A_c(z_3)$. Then the integrals over $z_1$ and $z_2$ can be done giving $A(z_3)$. So the result is

$$\frac{1}{V_c} \int dz_3 A(z_3) A_c(z_3). \tag{12}$$

If the cluster is small compared to the distance from its center to the observer, $A(z_3)$ is essentially constant over the region in which $A_c(z_3)$ is none zero. In this case $A(z_3) \simeq A(xD_s)$ and can be taken out of the integral. The integral is then just $V_c$ so equation 10 is verified. In the present situation the clusters have dimensions of parsecs and the distances to them are on the order of tens of kiloparsecs so this approximation should be valid. This result has also been verified numerically as mentioned in section 3.